\documentclass[letterpaper]{article} 
\usepackage{aaai25}  
\usepackage{times}  
\usepackage{helvet}  
\usepackage{courier}  
\usepackage[hyphens]{url}  
\usepackage{graphicx} 
\usepackage{natbib}  
\usepackage{caption} 
\usepackage{algorithm}
\usepackage{algorithmic}
\usepackage{amsthm}

\usepackage{subfigure}
\usepackage{array}
\usepackage{booktabs}
\usepackage{multirow}

\newcommand{\reffig}[1]{Figure\@~\ref{fig:#1}}

\usepackage{newfloat}
\usepackage{listings}
\usepackage{physics}

\DeclareCaptionStyle{ruled}{labelfont=normalfont,labelsep=colon,strut=off} 
\floatstyle{ruled}
\newfloat{listing}{tb}{lst}{}
\floatname{listing}{Listing}
\title{Q-MAML: Quantum Model-Agnostic Meta-Learning for Variational Quantum Algorithms}
\author{
    Junyong Lee\textsuperscript{\rm 1}\equalcontrib,
    JeiHee Cho\textsuperscript{\rm 2}\equalcontrib,
    Shiho Kim\textsuperscript{\rm 2}
}
\affiliations{
    \textsuperscript{\rm 1} BK21 Graduate Program in Intelligent Semiconductor Technology, Yonsei University, Korea\\
    \textsuperscript{\rm 2} Yonsei University, Korea\\
    jjunilee@yonsei.ac.kr, jeiheec@gmail.com, shiho@yonsei.ac.kr
}

\begin{document}

\maketitle

\begin{abstract}
In the Noisy Intermediate-Scale Quantum (NISQ) era, using variational quantum algorithms (VQAs) to solve optimization problems has become a key application. However, these algorithms face significant challenges, such as choosing an effective initial set of parameters and the limited quantum processing time that restricts the number of optimization iterations.
In this study, we introduce a new framework for optimizing parameterized quantum circuits (PQCs) that employs a classical optimizer, inspired by Model-Agnostic Meta-Learning (MAML) technique. This approach aim to achieve better parameter initialization that ensures fast convergence.
Our framework features a classical neural network, called \emph{Learner}, which interacts with a PQC using the output of \emph{Learner} as an initial parameter. During the pre-training phase, \emph{Learner} is trained with a meta-objective based on the quantum circuit cost function. In the adaptation phase, the framework requires only a few PQC updates to converge to a more accurate value, while the learner remains unchanged. 
This method is highly adaptable and is effectively extended to various Hamiltonian optimization problems. We validate our approach through experiments, including distribution function mapping and optimization of the Heisenberg XYZ Hamiltonian. 
The result implies that the \emph{Learner} successfully estimates initial parameters that generalize across the problem space, enabling fast adaptation.

\end{abstract}

%

\section{Introduction} \label{sec:intro}
\begin{figure}[t]
    \centering
    \includegraphics[width=\linewidth]{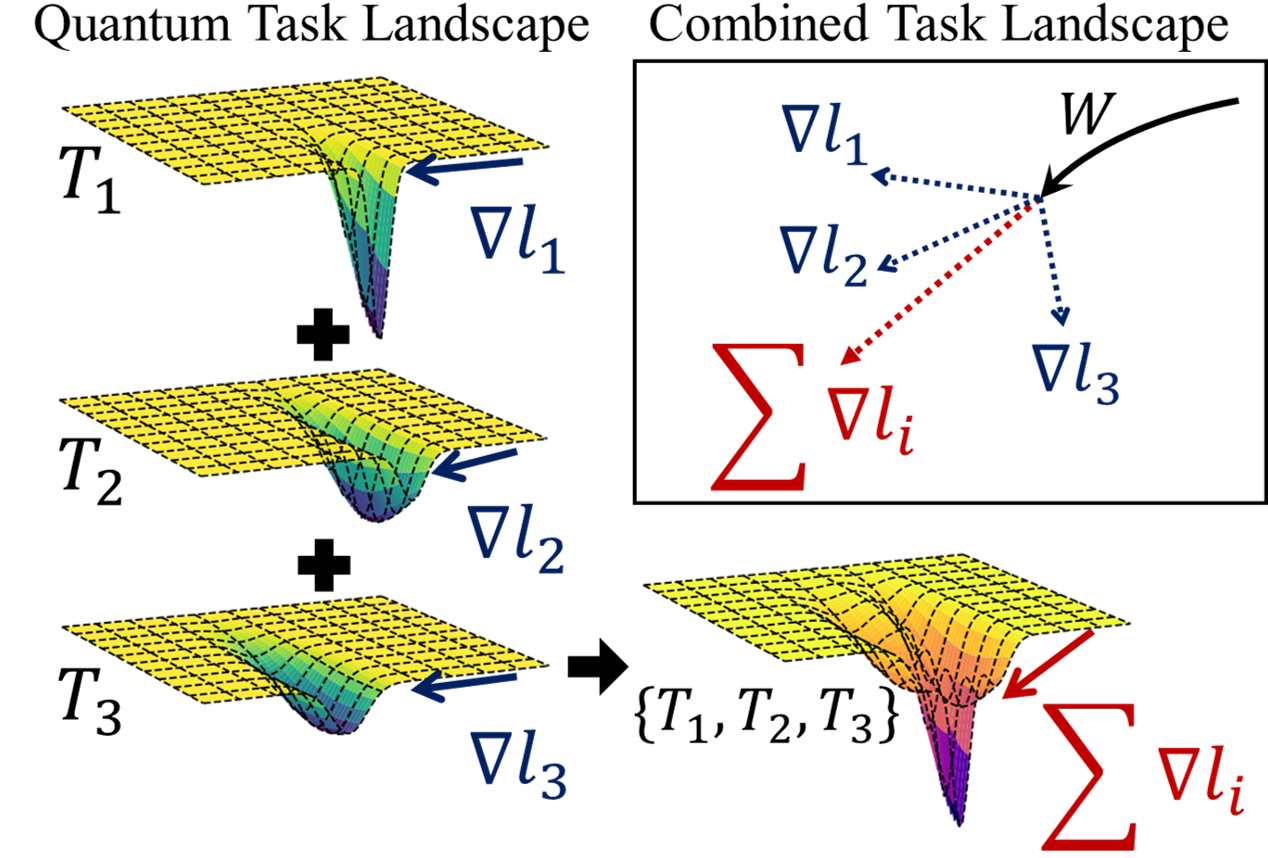}
    \caption{Illustration of the multiple individual optimization landscapes ($\nabla l_1, \nabla l_2, \nabla l_3)$ combine to form a single landscape ($\sum \nabla l_i$). The model trained using the combined landscape can capture the general properties of various tasks, which results in the improvement of convergence in the optimization as in the MAML.} 
    \label{fig:optimizaton}
\end{figure}

In the past decade, quantum computing has seen remarkable advancements, leading to the development of variational quantum algorithms (VQAs) that utilize parameterized quantum circuits (PQC), where gradient-based optimization methods are applied to gain optimal parameters~\cite{cerezo2021variational}. The popularity of this paradigm is due to its flexibility, widening the area of quantum applications and approaches~\cite{havlivcek2019supervised,huang2021experimental}. However, these methods face significant challenges due to the unfavorable characteristics of the optimization landscape.

One example is the Barren Plateau (BP), which occurs when the gradient of the cost function becomes exponentially small, caused by a flat optimization landscape~\cite{mcclean2018barren,wang2021noise,holmes2022connecting}. Lack of meaningful gradient value leads to failure of the learning, resulting in slow or non-convergence of the optimization process. BP is especially problematic as it scales unfavorably with the size of the quantum system, making it a crucial problem to solve for future advancement and usage of quantum computing~\cite{ragone2023unified}. 


\begin{figure*}
    \centering
    \includegraphics[width=0.8\linewidth]{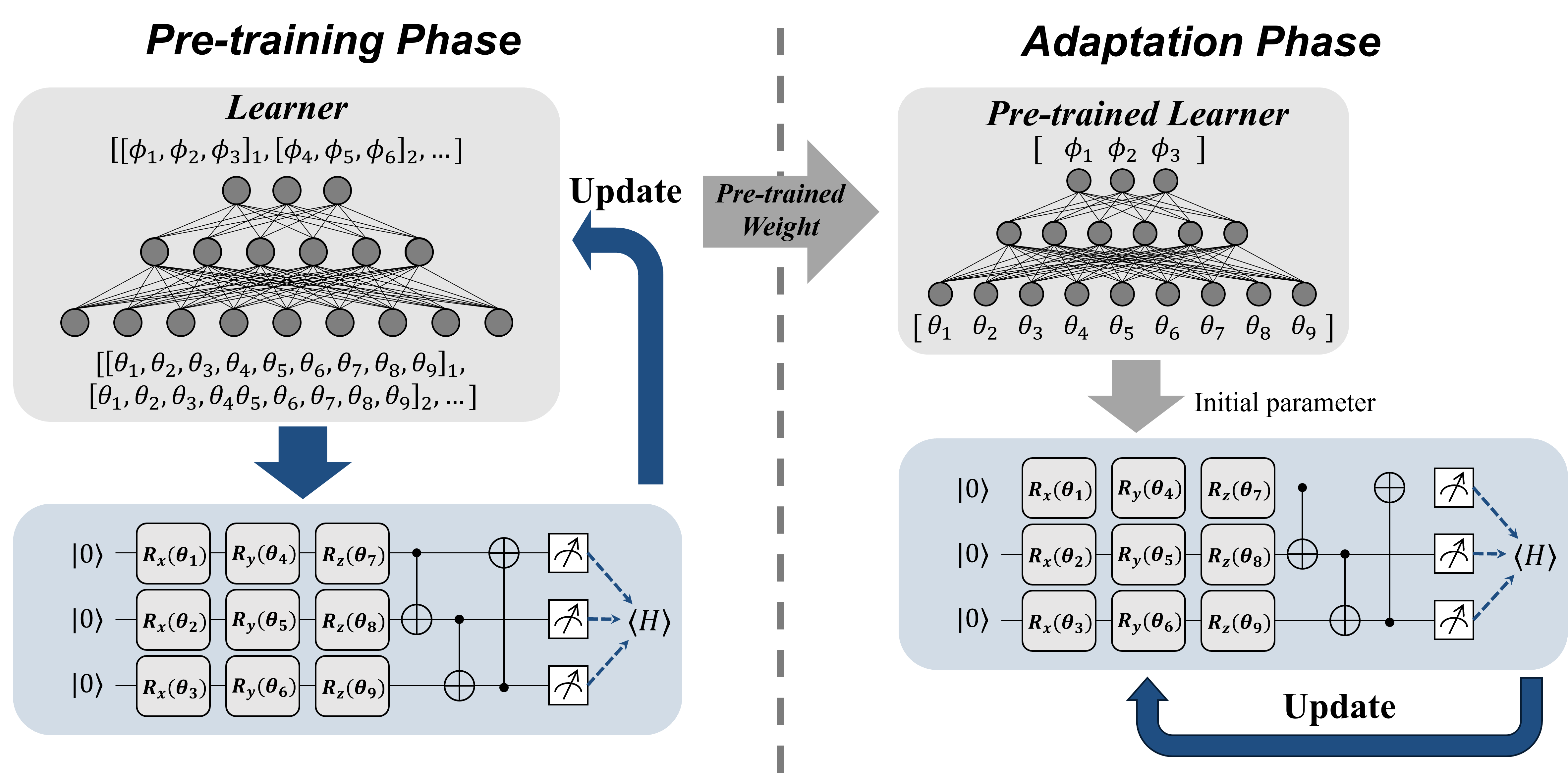}
    \caption{The overall learning path of Q-MAML. The pre-training phase includes an update of the \emph{Learner} while the adaptation phase trains PQC on specific tasks with parameter initialized using pre-trained \emph{Learner}.}
    \label{fig:framework}
\end{figure*}
Even with techniques designed to overcome the challenges posed by the optimization landscape, the iterative nature of quantum algorithms demands significant quantum computing resources. This makes the optimization process both time-consuming and costly, which can also be a burden for near-term quantum hardware. Therefore, achieving rapid convergence becomes essential to minimize computational expenses. A key strategy to enhance the performance and efficiency of PQC is identifying good initial parameters~\cite{friedrich2022avoiding}. Properly selected initial parameters can help mitigate the BP, facilitate more effective optimization, and ultimately improve the algorithms' ability to find meaningful solutions.

\citet{zhang2022escaping} suggest a Gaussian distribution-based parameter sampling method that can increase the expectation of gradient norm to reduce the likelihood of encountering BP.
Similarly, the reduced-domain parameter initialization technique shows promise in enhancing trainability~\cite{wang2023trainability}. 
However, these approaches limit the circuit design and therefore can not completely resolve the problem in more complex circuits or diverse VQA tasks. 
Additionally, they often overlook the specific characteristics of the Hamiltonian governing the quantum system, which can result in suboptimal performance in certain scenarios.


To tackle the challenges, we propose Q-MAML, a quantum-classical hybrid framework that uses the classical neural network, \emph{Learner}, to provide initial parameters for a PQC on optimization of various Hamiltonian. By leveraging Model-Agnostic Meta-Learning (MAML), \emph{Learner} can predict helpful initial parameters on various tasks, and therefore enable fast convergence of the PQC. We empirically confirm that our approach not only addresses the challenges posed by barren plateaus but also paves the way for more efficient and scalable quantum optimization solutions.

\section{Related Works} \label{sec:rel}
\subsubsection{Variational Quantum Algorithms (VQA)}
VQAs have emerged as a pivotal approach in leveraging near-term quantum devices to solve complex optimization problems~\cite{cerezo2021variational}. These algorithms combine PQCs with classical optimization techniques, making them versatile for a wide range of applications, including quantum chemistry, machine learning, and combinatorial optimization~\cite{lubasch2020variational, beckey2022variational, amaro2022case, jaksch2023variational}. VQAs are particularly appealing in that they are the leading candidate for achieving a quantum advantage in the NISQ era.

One of the most promising VQAs is Variational Quantum Eigensolvers (VQEs), which are designed for solving quantum chemistry and material science problems~\cite{peruzzo2014variational}. The goal of VQE is to find the ground state energy of a Hamiltonian $H$ by minimizing the cost function, defined as follows:
\begin{equation}
    E(\theta)=\bra{\psi(\theta)}H\ket{\psi(\theta)}
\end{equation}
where $\psi$ is quantum state and $\theta$ is parameter of PQC.

Despite their potential, VQAs face significant challenges, particularly in terms of trainability and scalability. One of the primary issues is the BP phenomenon~\cite{mcclean2018barren}, where the gradient of the cost function becomes exponentially small, hindering the optimization process. These limitations underscore the importance of developing more robust initialization techniques that incorporate Hamiltonian-specific information, as our research proposes. By aligning initial parameters with the intrinsic properties of the Hamiltonian, our approach aims to enhance the trainability and convergence of VQAs, offering a more effective solution to the challenges faced by current methods.

There are several approaches to enhance trainability with proper initialization of circuits~\cite{zhang2022escaping,metavqe}, however, they require specific circuit architecture or lack of explanation of generalization.

\subsubsection{Model-Agnostic Meta-Learning (MAML)}
MAML is a highly influential approach in meta-learning, designed to allow models to adapt to new tasks with minimal data. Introduced by~\citet{finn2017model}, MAML operates by optimizing a model's parameters such that only a few gradient updates are required for fast adaptation to new tasks. This makes MAML particularly effective for few-shot learning, where the goal is to generalize from limited examples.

MAML's versatility lies in its model-agnostic characteristic, meaning it can be applied across a wide range of tasks, including supervised learning, reinforcement learning, and regression. By focusing on creating a universal initialization that performs well across different tasks, MAML has become a cornerstone in meta-learning research, influencing a broad range of subsequent studies and applications~\cite{wang2022global}.

Inspired by the advantages of MAML, we aim to provide a universal initialization point for various VQE tasks. In VQE, the Hamiltonian is considered a single task, and the task space is defined as the space of optimization in the D-dimensional parameter space, where D is the total number of parameters being optimized. We apply the MAML method to train the \emph{Learner} by constructing this task space with different Hamiltonians. With Q-MAML, the \emph{Learner} gains the ability to guess good initial parameters for optimizing various Hamiltonians.
\section{Method} \label{sec:main}


In this section, we illustrate detailed procedures of Q-MAML that consist of two key phases: the pre-training phase and the adaptation phase.

In the pre-training phase, \emph{Learner} explores the combined task space (see~\reffig{optimizaton}) to learn a robust initialization point of a quantum circuit that can quickly adapt to various quantum tasks. Similar to MAML, the \emph{Learner} is generalized across different Hamiltonian tasks, capturing essential features and common patterns within this domain.
In the adaptation phase, a specific Hamiltonian is selected and the PQC is fine-tuned to minimize the cost/loss function with a classical optimizer. During this stage, the PQC refines its weights to optimize cost on the target task as in VQE.

This dual-phase approach allows faster convergence and better optimization under the constraints of limited quantum processing time. The overview of Q-MAML is provided in~\reffig{framework}.

\begin{algorithm}[tb!]
\caption{MAML for Pre-training Phase}
\label{alg:pretrain}
\textbf{Parameter} $\alpha$: learning rate
\begin{algorithmic}[1]
\REQUIRE $p(\mathcal{T})$: distribution of tasks
\REQUIRE classical optimizer $\mathcal{O}$, cost function $l_{T}(\boldsymbol{\theta})$
\STATE randomly initialize $W$
    \WHILE{not done}
    \STATE Sample batch of task $T_i \sim p(\mathcal{T})$
        \FORALL{$T_i$}
            \STATE Get $\theta$ from \emph{Learner} : $\theta = h_W(\phi_i)$
            \STATE Prepare quantum state $\ket{\psi(\boldsymbol{\theta})} = g(\boldsymbol{\theta}) \ket{0}$
            \STATE Evaluate cost function $l_{T_i}(\boldsymbol{\theta}) = \langle \psi(\boldsymbol{\theta}_k) | \hat{H} | \psi(\boldsymbol{\theta}_k) \rangle$
            \STATE Calculate the gradient $\nabla_W l_{T_i}(g(h_W(\phi_i)))$
        \ENDFOR
        \STATE Update parameter \\
        $W \leftarrow W + \alpha\nabla_{W}\sum_{T_i}l_{T_i}(g(h_W(\phi))$
    \ENDWHILE
\end{algorithmic}
\end{algorithm}

\begin{algorithm}[tb]
\caption{Adaptation to Single Task}
\label{alg:adaptation}
\textbf{Input}: $T^*$: Target Task (Hamiltonian)\\
\textbf{Parameter}: $\alpha$: learning rate
\begin{algorithmic}[1]
\REQUIRE Parameterized Quantum Circuit $g(\boldsymbol{\theta})$
\REQUIRE classical optimizer $\mathcal{O}$, cost function $l_{T^*}(\boldsymbol{\theta})$
\REQUIRE: $h_{W^*}$: pre-trained \emph{Learner}
\STATE Get initial parameters $\theta = h_{W^*}(\psi)$
\WHILE{not done}
    \STATE Prepare quantum state $\ket{\psi(\boldsymbol{\theta})} = g(\boldsymbol{\theta}) \ket{0}$
    \STATE Evaluate cost function $l_{T^*}(\boldsymbol{\theta}_k) = \langle \psi(\boldsymbol{\theta}_k) | \hat{H} | \psi(\boldsymbol{\theta}_k) \rangle$
    \STATE Calculate the gradient $\nabla_{\boldsymbol{\theta}} l_{T^*}(\boldsymbol{\theta}_k)$
    \STATE Update parameter \\
    $\theta \leftarrow \theta + \alpha \nabla_{\boldsymbol{\theta}} l_{T^*}(\boldsymbol{\theta}_k)$
\ENDWHILE\\

\end{algorithmic}
\textbf{Output} Final optimization $\boldsymbol{\theta}^*$, minimal cost $l_{T^*}(\boldsymbol{\theta}^*)$
\end{algorithm}

\subsection{Pre-training Phase}
The pre-training phase focuses on determining an effective initial set of parameters for the PQC, denoted as $\theta$. During this phase, a classical neural network (\emph{Learner}) is employed to approximate the initial parameters by minimizing a meta-objective function based on the quantum circuit's cost function (Equation~\ref{eq:objective}). The mathematical background of Hamiltonian task space is provided in the Appendix. The detailed algorithm is provided in Algorithm~\ref{alg:pretrain}.

The \emph{Learner}'s input is the Hamiltonian task vector ($\phi$), and the \emph{Learner}'s output is assigned as a parameter of the PQC ($\theta$), and the observable of the PQC ($\langle\boldsymbol{O}\rangle$) is used for the cost calculation. In this structure, the Meta-Learning Objective is as follows:
\begin{equation}
    \underset{W}{\mathrm{argmin}} \sum\limits_{T_i\sim p(\mathcal{T})} l_{T_i}(g(h_{W}(\phi_i)))
    \label{eq:objective}
\end{equation}
where $l_{T_i}$ is the cost function for task $T_i$.
Given the known fact that the combination of cost functions is feasible in PQC training~\cite{nemkov2024barren}, we have taken advantage of this by differentiating the \emph{Learner}'s weight concerning the combined cost function, instead of computing and directly updating $\theta$ using its Hessian as in MAML. This approach allows for more efficient and scalable optimization, aligning the parameter updates with the underlying quantum task characteristics.
The \emph{Learner}'s weight is trained via backpropagation of the gradients, where the chain rule is still applicable. With generated observable $\langle\boldsymbol{O}\rangle$, the gradient to update $W$ can be calculated as follows:
\begin{equation}
    \frac{dl_T}{dW} = \frac{dl_T}{d\langle\boldsymbol{O}\rangle} \frac{d\langle\boldsymbol{O}\rangle}{d\theta} \frac{d\theta}{dW}
\end{equation}
where $\frac{dl_T}{d\langle\boldsymbol{O}\rangle}$ is differentiable as cost functions are typically defined to be differentiable. Additionally, $\frac{d\langle\boldsymbol{O}\rangle}{d\theta}$ is differentiable as is confirmed through training in VQA and \citet{nemkov2024barren}, and $\frac{d\theta}{dW}$ represents standard differentiation in neural network training.
The \emph{Learner} is trained using a variety of Hamiltonians to ensure that the initial parameters it generates are generalizable and serve as a good starting point for different quantum tasks. This phase aims to reduce the burden on the quantum processor by providing a well-informed initialization, which reduces the number of iterations required during the adaptation phase. 



\subsection{Adaptation Phase}
Following the pre-training phase, the adaptation phase involves fine-tuning the PQC using the initial parameters $\theta$ estimated by the \emph{Learner}. Unlike traditional VQAs, where parameter optimization is performed from scratch, our method utilizes the pre-learned parameters as a starting point. Other components such as cost and classical optimizer are the same as traditional VQAs. In this phase, only the PQC is updated, and no further adjustments are made to the \emph{Learner}. This focused adaptation ensures that the quantum circuit can adapt to the specific problem at hand with minimal quantum resource usage. Detailed algorithm is provided in Algorithm~\ref{alg:adaptation}.


\begin{figure*}[hbt!]
    \centering
     \subfigure[Heisenberg 12 qubits]{\includegraphics[width=0.47\columnwidth]{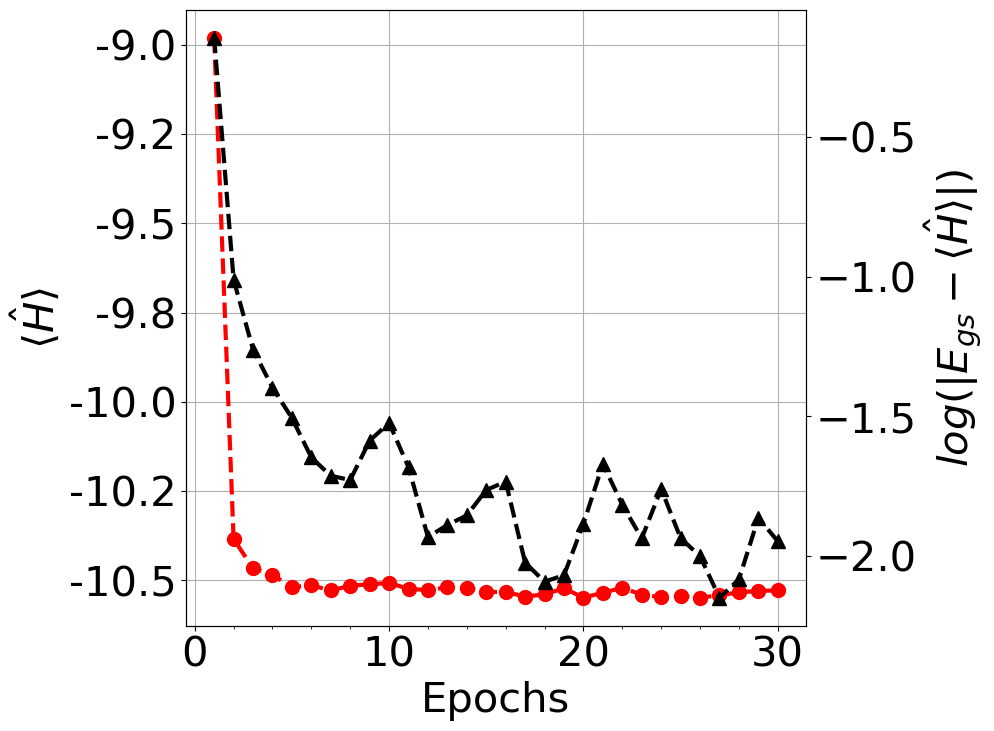} \label{fig:xyz_12_traj}}
     \subfigure[Heisenberg 20 qubits]{\includegraphics[width=0.47\columnwidth]{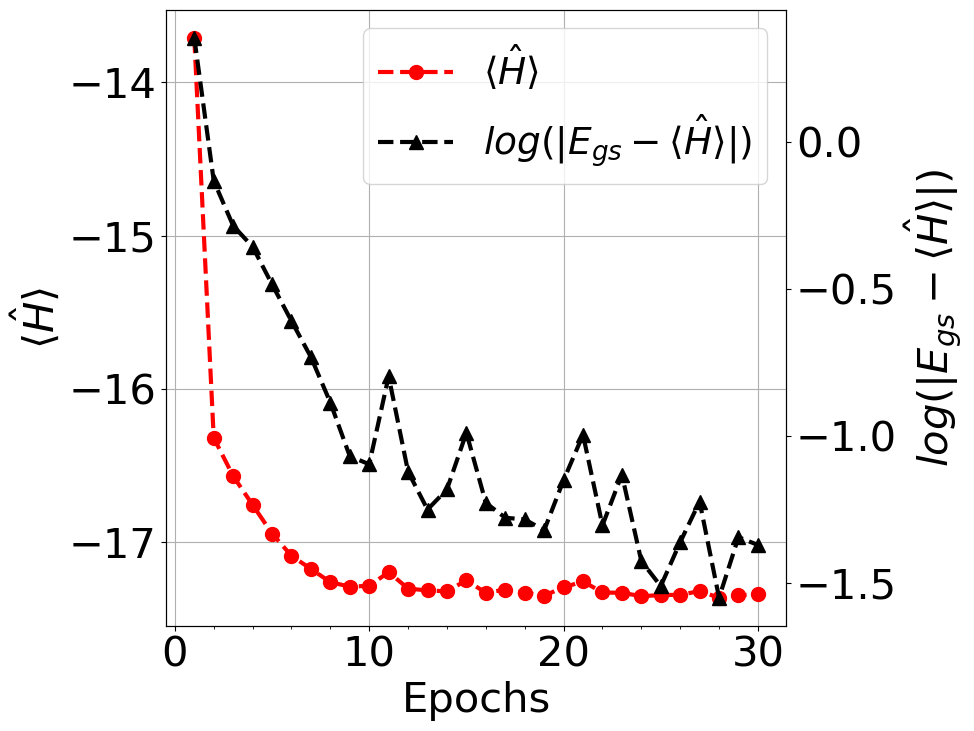} \label{fig:xyz_20_traj}}
     \subfigure[Molecule 10 qubits]{\includegraphics[width=0.47\columnwidth]{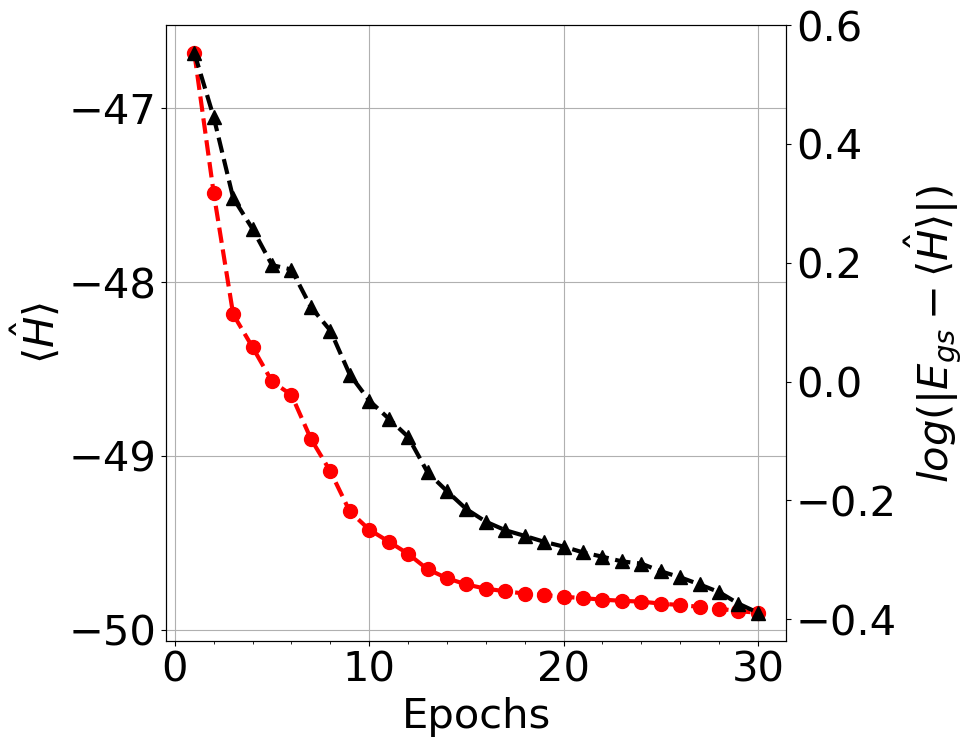} \label{fig:molecule_10}}
     \subfigure[Molecule 14 qubits]{\includegraphics[width=0.45\columnwidth]{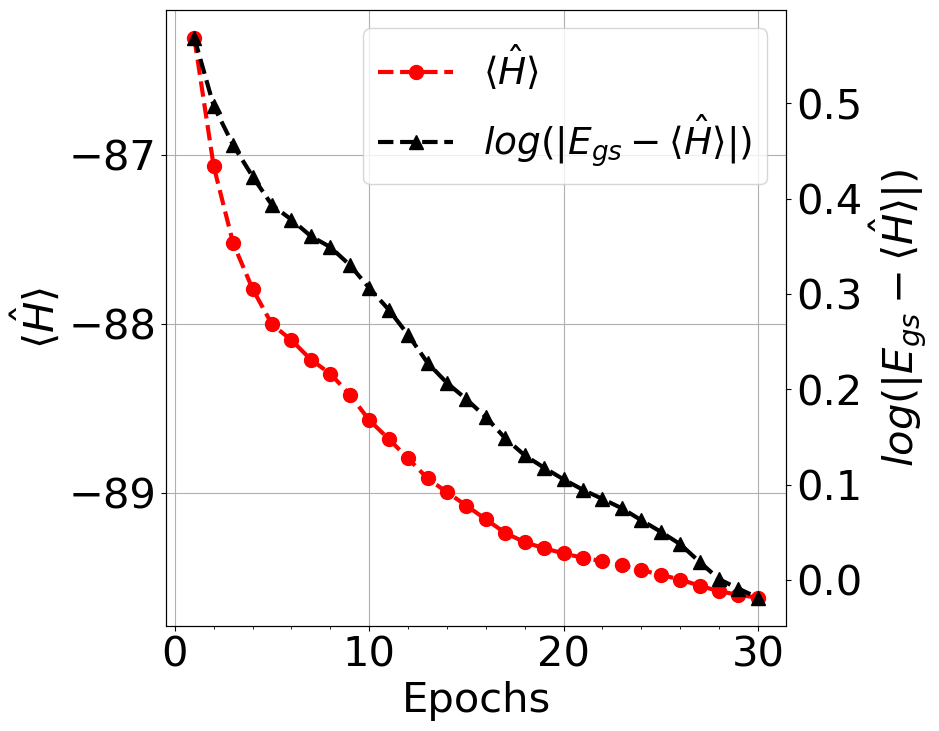} \label{fig:molecule_14}}
    \caption{Training trajectory of \emph{Learner} with different Hamiltonian dataset.}
    \label{fig:learner_traj}
\end{figure*}
\begin{figure*}
    \centering
     \subfigure[Heisenberg 12 qubits]{\includegraphics[width=0.45\columnwidth]{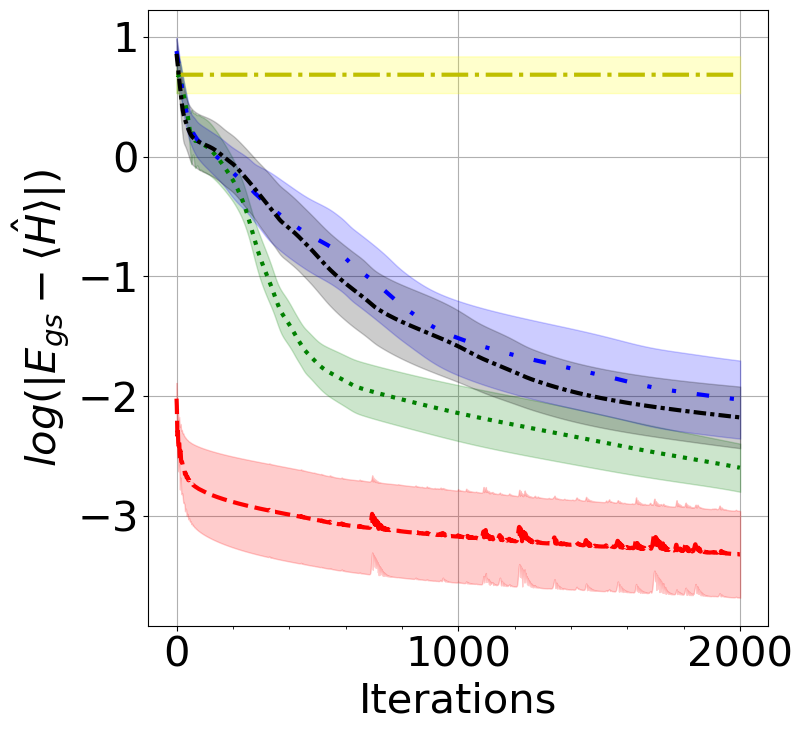} \label{fig:xyz_12_adapt}}
     \subfigure[Heisenberg 20 qubis]{\includegraphics[width=0.45\columnwidth]{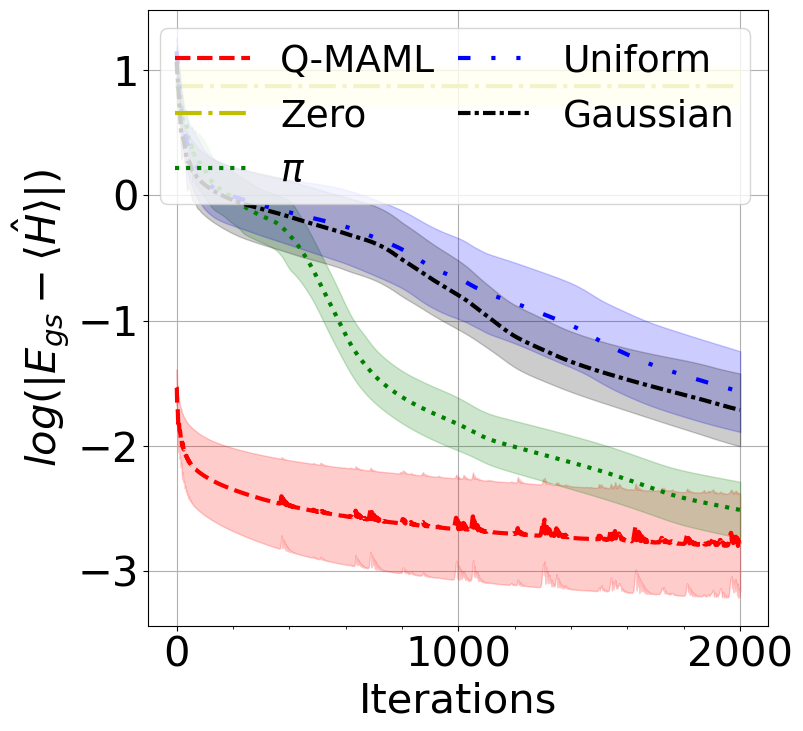} \label{fig:xyz_20_adapt}}
     \subfigure[Molecule 10 qubits]{\includegraphics[width=0.47\columnwidth]{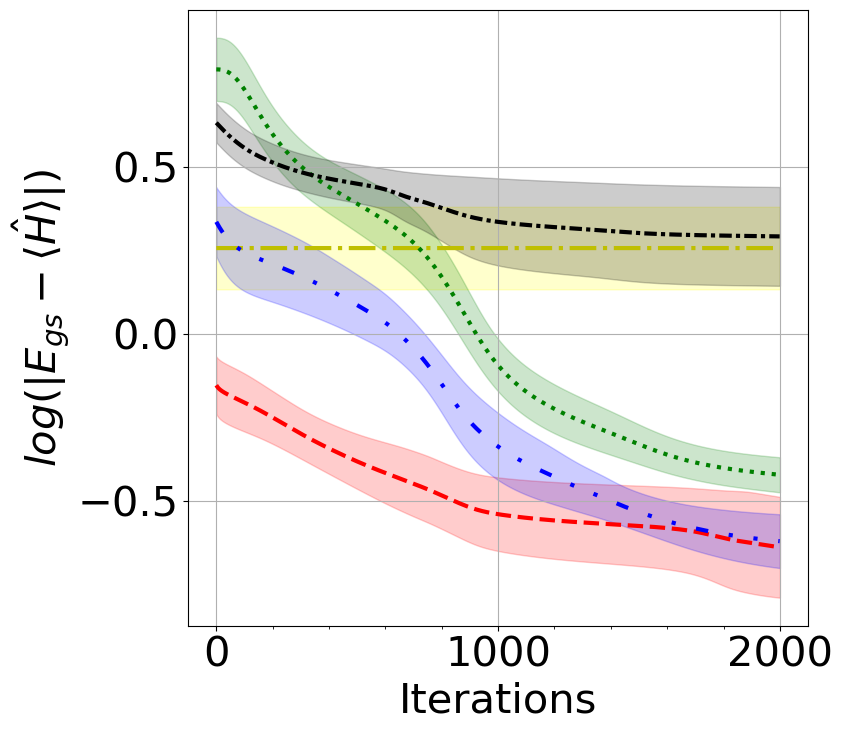} \label{fig:molecule_10_adap}}
     \subfigure[Molecule 14 qubits]{\includegraphics[width=0.47\columnwidth]{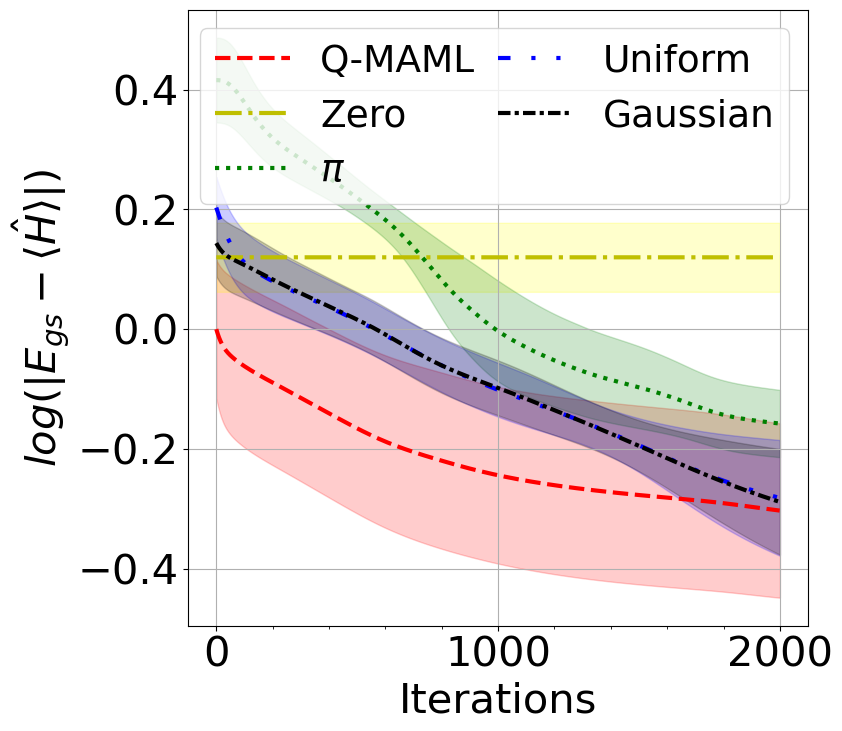} \label{fig:molecule_14_adap}}
    \caption{Training trajectory of PQC in the adaptation phase. The average gap value is plotted using 16 and 6 different Hamiltonian for Heisenberg and Molecule, respectively. The shaded regions indicate one standard deviation ($[\mu-\sigma/2, \mu+\sigma/2]$).}
    \label{fig:pqc_traj}
\end{figure*}

\section{Experiments and Results} \label{sec:exp}

We conduct experiments on two different applications: Heisenberg XYZ Hamiltonian, and Molecule Hamiltonian. To show the expandability of the framework, we perform additional application, and distribution function embedding and provide results in the Appendix. 

Throughout the experiments, the \emph{Learner} is trained for 30 epochs using Adam optimizer with a learning rate of 0.001. For the adaptation phase, PQC is 
trained for 2000 iterations by Adam optimizer with 0.001 as a learning rate. We use four different initialization methods that are commonly used: Zero initialization, $\pi$ initialization, reduced-domain Uniform initialization~\cite{wang2023trainability}, and Gaussian initialization~\cite{zhang2022escaping}. Zero and $\pi$ initialization method use zero and $\pi$ for all parameters. Parameters of reduced-domain Uniform initialization is sampled from $\mathcal{U}[-\alpha\pi, \alpha\pi]$ where $\alpha$ is set as $0.05$ to ensure sufficiently small values are sampled, and Gaussian initialization from $\mathcal{N}(0,\gamma^2)$ and set $\gamma^2=\frac{1}{4S(L+2)}$ where $S$ is the number of non-identity Pauli operators in observable and $L$ is the number of layers in the circuit that follows the setting of~\citet{zhang2022escaping}. 


\subsection{Heisenberg XYZ Hamiltonian}
The Heisenberg XYZ Hamiltonian is a quantum mechanical model that describes interactions between spins on a lattice. It is an extension of the Heisenberg model, which is used to understand the magnetic properties of materials at the quantum level. The XYZ Hamiltonian specifically accounts for anisotropic interactions in three different spatial directions (X, Y, and Z), meaning that the interaction strengths along these directions can differ. 
Heisenberg XYZ Hamiltonian is defined via three parameters, $J_x$, $J_y$, and $J_z$ and can be obtained by calculating following equation:
\begin{equation}
\begin{aligned}
    &\hat{H}_{XYZ} \\
    &= - \sum\limits_{n=1}^{N-1} (J_x\sigma^1_n \sigma^1_{n+1} + J_y\sigma^2_n \sigma^2_{n+1} + J_z\sigma^3_n \sigma^3_{n+1} +h \sigma_n^3) 
    \label{eq:heisenberg}
    \end{aligned}
\end{equation}
where $\sigma^1, \sigma^2,$ and $\sigma^3$ represent the Pauli X, Y, and Z matrices, respectively, and $h$ is the effect of the external magnetic field where we set as 0.

\subsubsection{Task Space Definition}
Heisenberg XYZ Task Space consists of vectors representing each Hamiltonian. As Hamiltonian varies by $J_i$ (Equation~\ref{eq:heisenberg}), each task (Hamiltonian) can be defined by $J^i = [J_x^i, J_y^i, J_z^i]$ where $i$ indicates task number. We randomly sample three values within the range~$[-3,3]$ with an interval of 0.1 to construct $J^i$.

\subsubsection{Experiment Setting}
The \emph{Learner} is constructed using one input layer with a size of 3, two hidden layers with 256 neurons, and an output layer with a size of the number of parameters in PQC. We vary the number of qubits by 12 and 20 to check the performance of Q-MAML on a small and large number of qubits. Experiments using 10, 14, 16, and 18 qubits are provided in the Appendix along with the design of the ansatz. The ansatz of PQC consists of blocks that have IsingXX, IsingYY, and IsingZZ gates. We randomly select 16 samples to check the adaptation performance of Q-MAML.

\subsection{Molecule Hamiltonian}
The Molecule Hamiltonian is the crucial point for solving the Schrödinger equation for a molecule, which yields important information about the molecule's electronic structure, bond lengths, bond angles, and other physical and chemical properties. For $M$ spin molecular orbital (MO), the basis is denoted as $\ket{n_0,n_1,...,n_{M-1}}$. Each basis can take 0 or 1 which indicates the electron occupancy of the corresponding orbital. It can be written as:
\begin{equation}
    \hat{H} = \sum\limits_{p,q}h_{p,q}c_p^{\dagger}c_p + \frac{1}{2}\sum\limits_{p,q,r,s}c_p^{\dagger}c_q^{\dagger}c_r c_s,
\end{equation}
$c^{\dagger}$ and $c$ are the electron creation and annihilation operators, respectively, while the coefficients $h_{pq}$ and $h_{pqrs}$ correspond to the one and two-electron Coulomb integrals, estimated by the Hartree-Fock orbitals. Using $M$ qubits to encode the $M$ numbers of basis.
Through the Jordan-Wigner transformation, fermionic operators are mapped onto qubit operators, expressed as a tensor product of Pauli matrices ${I, X, Y, Z}$ action on the qubits.
\begin{equation}
    \hat{H} = \sum\limits_{j}C_j\otimes_i\sigma_i^{(j)}
\end{equation}
where $\sigma \in {I, X, Y, Z}$. 

\subsubsection{Task Space Definition}
Molecule Hamiltonian Task Space is constructed with the vector $C$, which is the coefficient of the molecule Hamiltonian. It should be noted that each task can have a different length.

\subsubsection{Experiment Setting}
We apply the same \emph{Learner} architecture as in Heisenberg XYZ Hamiltonian. The only difference is the input size of the \emph{Learner}, where the max length of $C$ is used in this experiment. If $C$ is shorter than the max length, the remaining values are zero-padded. The PQC consists of 7 layers of StronglyEntanglingLayer~\cite{schuld2020circuit}, with an example of the circuit provided in the Appendix.
We randomly select 6 samples (which is 10\% of the total dataset) to evaluate the performance of adaptation.

\subsubsection{Dataset Generation}
We use Pennylane function \emph{qml.molecular\_hamiltonian} to generate a Hamiltonian of molecules with specific conditions. We give 12 distinct bond lengths to generate different Hamiltonian as many as possible. As the number of qubits required for Hamiltonian is related to the number of active orbitals, we set it as 5 and 7 to generate 10 and 14 qubits Hamiltonian. Detailed parameters to generate the data are provided in the Appendix.

\subsubsection{Results}
We first validate the learning of the \emph{Learner} on the task space. In~\reffig{learner_traj}, we can tell from the result that the \emph{Learner} successfully learned the pattern of the task space and achieved proper estimation resulting in a small gap between the ground state energy concerning the dataset and the number of qubits.

With pre-trained \emph{Learner}, the initial parameter for various tasks is generated for the adaptation phase. In~\reffig{pqc_traj}, we visualize the learning curve of PQC to assess its convergence speed and performance. The gap between ground state energy calculated using exact diagonalization and expectation of PQC is measured. It should be noted that, unlike training of the \emph{Learner}, separate PQC is used and trained per task. The $\gamma^2$ for Gaussian initialization is set as $\frac{1}{176}$ where $S=2$ and $L=20$ as we use at most $20$ layers for experiments using Heisenberg dataset, and $\frac{1}{360}$, $\frac{1}{504}$ respectively for 10 and 14 qubit Molecule experiments. The result implies that when initialized with Q-MAML, PQC can converge faster and therefore requires lower quantum resources compared to other methods. 

\section{Discussion} \label{sec:dis}
In this section, we delve into the key findings. Focusing on three primary of interest: the learning behavior of the \emph{Learner}, its impact on gradient norm characteristic, and the barren plateau phenomenon. First, we examine what the \emph{Learner} is actually learning during the pre-training phase, and how this knowledge translates into effective parameter initialization for the quantum circuit. Next, we analyze the influence of the \emph{Learner}'s initialization on the gradient norms throughout the optimization process, shedding light on its role in improving convergence. Finally, we address the ongoing challenge of barren plateaus, discussing how our approach mitigates this issue and comparing our results with existing strategies in the field.

\subsection{What the \emph{Learner} Discovers During Training?}
First, we define the Task Space $\mathcal{T}$ and hypothesize the relationship between this space and the quantum circuit parameter $\theta$. Q-MAML approach is designed to learn this relationship during the pre-training phase, aiming to produce better initial parameters for a given task. The key insight extracted from the following experiment is that the \emph{Learner} learns to capture the distinctive patterns across the task space, and therefore effectively tailor the initialization for each Hamiltonian.

\begin{figure*}[ht]
    \centering
     \subfigure[Heisenberg 12 qubits]{\includegraphics[width=0.45\columnwidth]{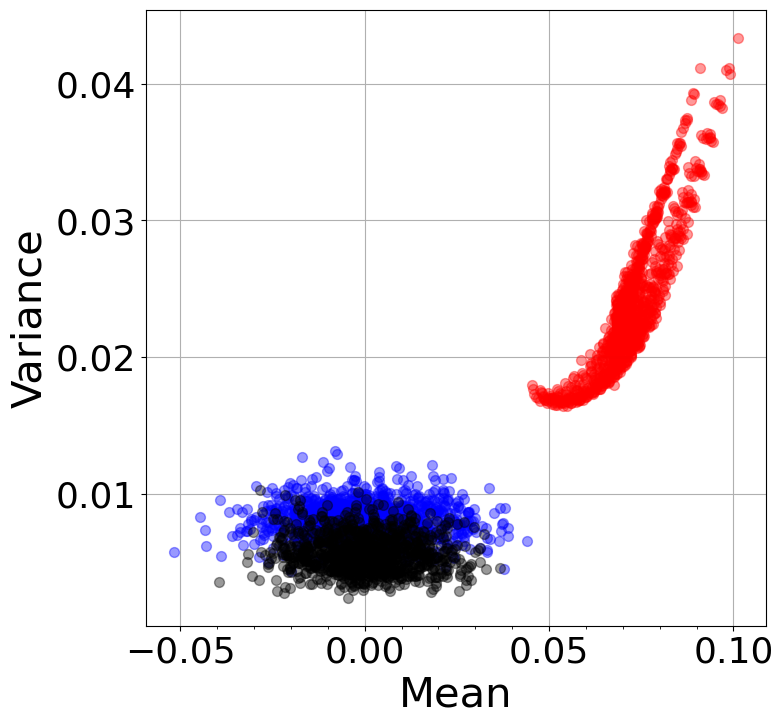} \label{fig:xyz_12_stat}}
     \subfigure[Heisenberg 20 qubits]{\includegraphics[width=0.45\columnwidth]{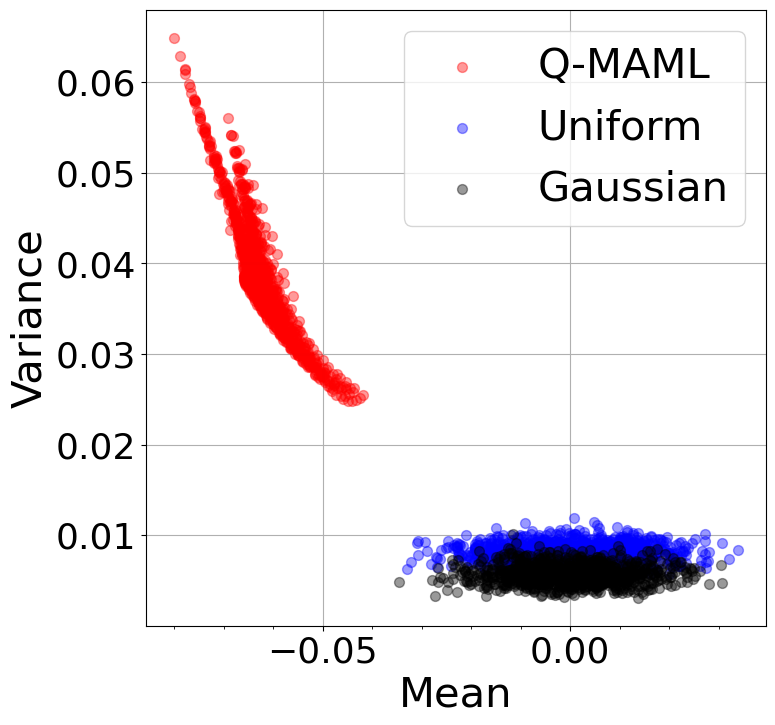} \label{fig:xyz_20_stat}}
     \subfigure[Molecule 10 qubits]{\includegraphics[width=0.45\columnwidth]{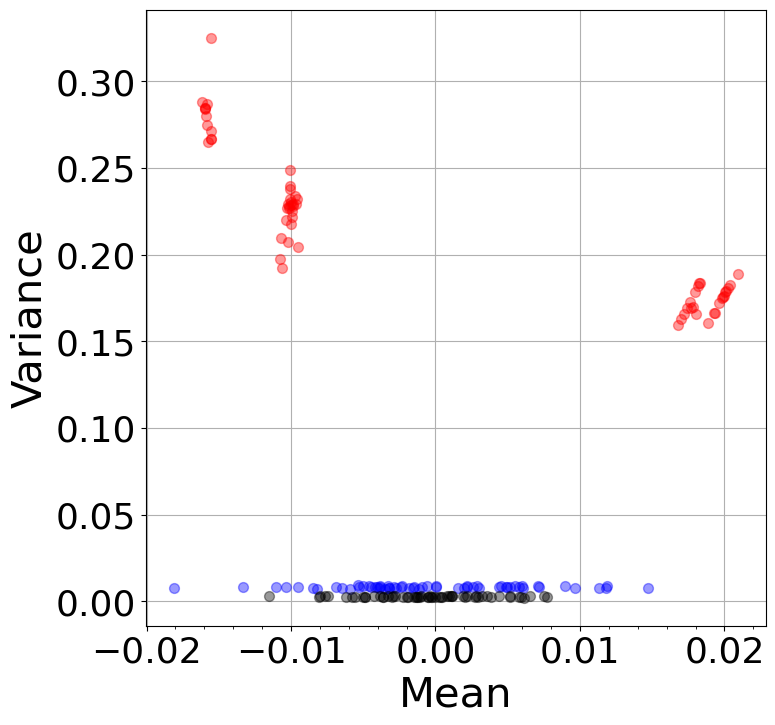} \label{fig:molecule_10_dist}}
     \subfigure[Molecule 14 qubits]{\includegraphics[width=0.45\columnwidth]{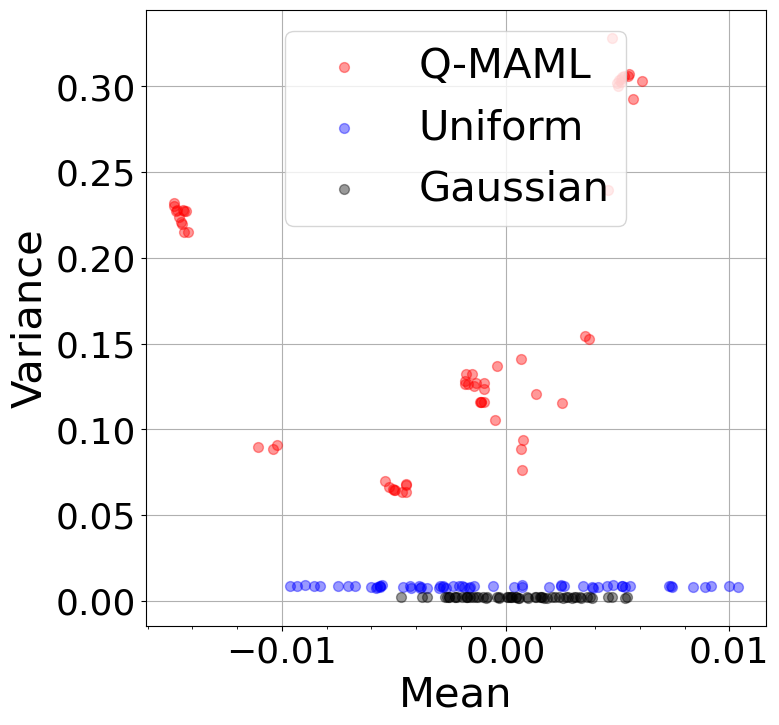} \label{fig:molecule_14_dist}}
    \caption{Statistics of the parameter initialized using Q-MAML, Uniform, and Gaussian.}
    \label{fig:gen_stat}
\end{figure*}

\begin{figure*}[ht]
    \centering
     \subfigure[Heisenberg 12 qubits]{\includegraphics[width=0.45\columnwidth]{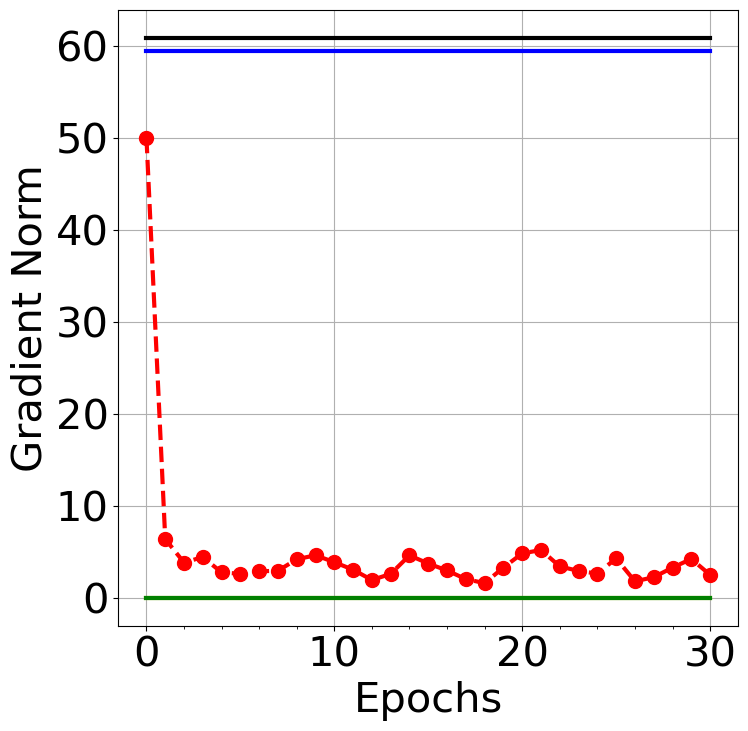} \label{fig:xyz_10_grad}}
         \subfigure[Heisenberg 20 qubits]{\includegraphics[width=0.47\columnwidth]{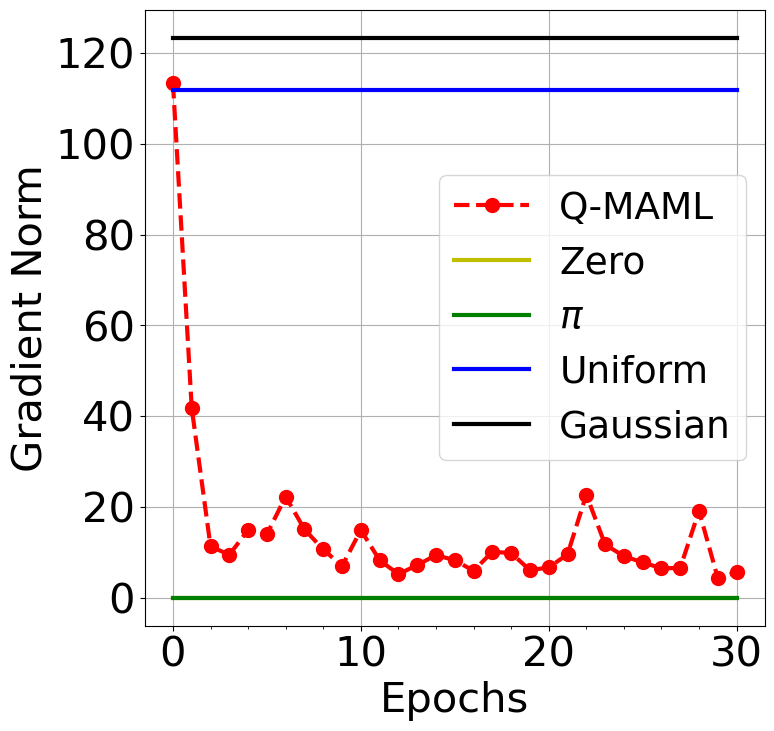} \label{fig:xyz_20_grad}}
     \subfigure[Molecule 10 qubits]{\includegraphics[width=0.45\columnwidth]{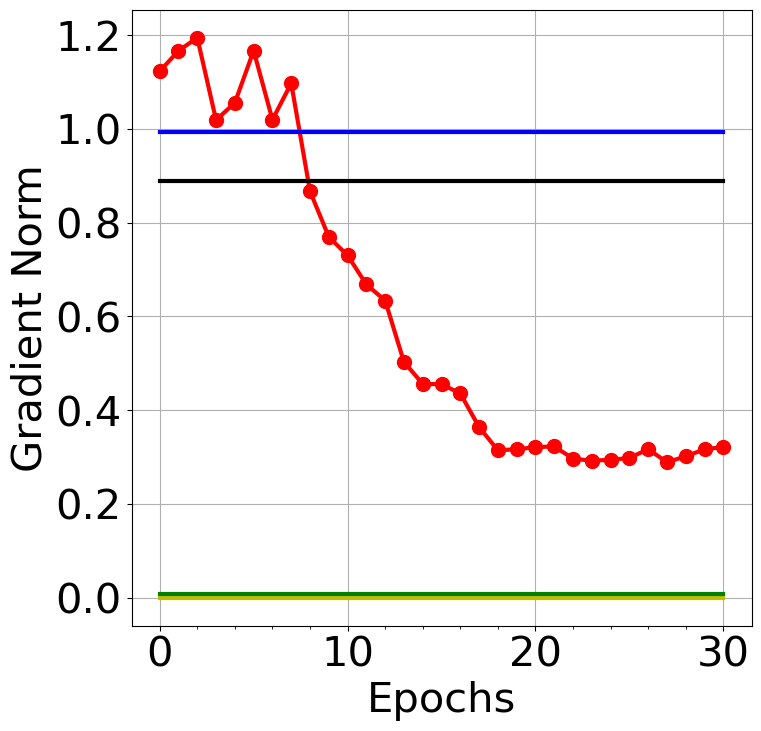} \label{fig:molecule_10_grad}}
         \subfigure[Molecule 14 qubits]{\includegraphics[width=0.47\columnwidth]{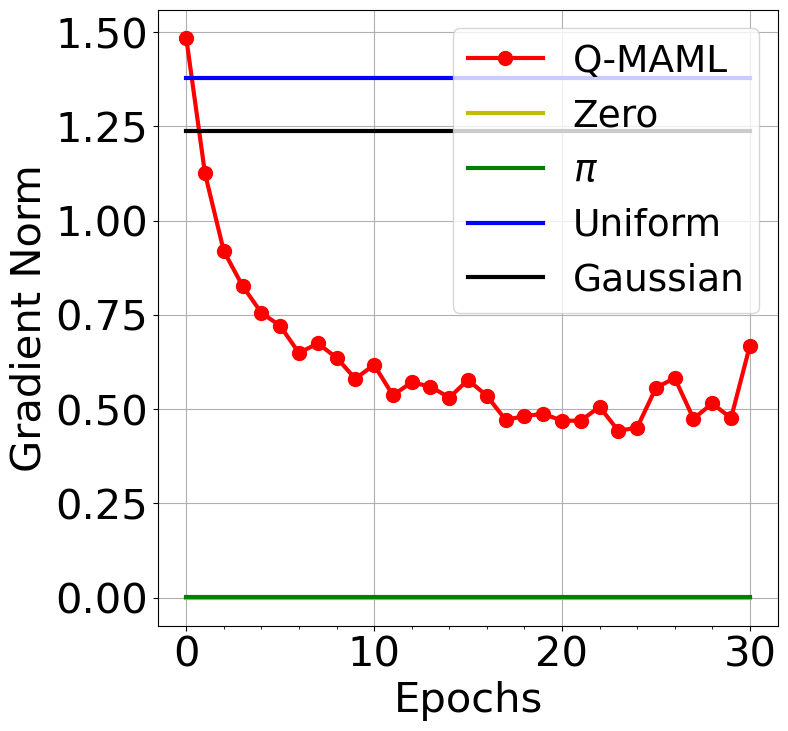} \label{fig:molecule_14_grad}}
    \caption{Trend of the PQC's gradient norm during pre-training phase}
    \label{fig:grad_curve}
\end{figure*}

Figure~\ref{fig:gen_stat} shows the statistical properties of the parameters initialized using Q-MAML, compared to Uniform and Gaussian initialization methods, across different Task Space and qubits configurations. In both figures, 
Q-MAML shows a broader distribution in terms of variance and mean. This indicates that the \emph{Learner} is identifying and leveraging specific patterns within each Task Space to generate diverse and task-specific initializations. These patterns are not present in the Uniform or Gaussian methods, which result in more constrained parameter distributions.

By adapting to the specific characteristics of Task Space, the Learner not only provides task-specific initialization but also significantly enhances the optimization process. This is demonstrated in the pre-training and adaption trajectories (Figure~\ref{fig:learner_traj},~\ref{fig:pqc_traj}), where Q-MAML consistently exhibits superior starting points compared to other methods. As Figure~\ref{fig:pqc_traj} illustrates, Q-MAML's initialized parameters display a broader and more effective distribution, which directly contributes to more efficient convergence. The \emph{Learner}’s ability to identify and incorporate distinct patterns across different Task Spaces results in these advantageous starting positions, highlighting the effectiveness of our approach and its potential to address the challenges posed by barren plateaus in VQAs.

\subsection{How the \emph{Learner} affect Gradient Norms?}

The norm of the gradient is a widely recognized tool for assessing the convergence behavior of VQAs. In many studies, tracking the gradient norm provides valuable insights into the optimization process and helps identify the factors that influence convergence rates. In this section, we explore the fast adaptation observed with Q-MAML by analyzing how the gradient norm evolves during the pre-training process. By examining the train trends in the gradient norm in different qubit configurations, as illustrated in Figure~\ref{fig:grad_curve}, we aim to understand the mechanisms that contribute to the enhanced convergence performance of our approach.

In classical neural networks, the norm of the gradient is a critical indicator of learning stability and convergence. A well-balanced gradient norm ensures stable learning, preventing issues such as "exploding gradients," where overly large gradients cause instability, or "vanishing gradients," where excessively small gradients hinder progress~\cite{hochreiter2001gradient, glorot2010understanding, pascanu2013difficulty}. By maintaining a moderate gradient norm, the learning process can make consistent and meaningful updates to the model parameters, leading to efficient and reliable convergence. This concept underscores the importance of monitoring gradient norms to achieve optimal performance during the gradient-based learning process, which applies to both neural network and PQC training.

To better predict where our initial point will be and how the model will converge, we measure the average gradient norm at the starting point of each epoch across the entire dataset. This approach allows us to track how the initial gradients evolve and gives us insight into the expected trajectory of the optimization process. In addition, it is one of the important evidence to predict the overall stability of the VQA optimization.

In \reffig{grad_curve}, we depict the trends of the gradient norm in different configurations. Overall, the gradient norm of Q-MAML converges to a specific value, which is \emph{moderate} compared to other approaches.
Although this observation needs further discussion, we can assume that the \emph{Learner} provides better initial $\theta$ for adaptation relative to other approaches with larger gradient norm. This implies that Q-MAML effectively mitigates issues such as exploding gradients, resulting in more efficient and reliable convergence against traditional initialization methods. An additional discussion on this topic is concluded in the following section, \emph{A New Perspective on Barren Plateaus}.

\subsection{A New Perspective on Barren Plateaus}
\begin{figure}[ht]
    \centering
     \subfigure[Heisenberg 20 qubits]{\includegraphics[width=0.45\columnwidth]{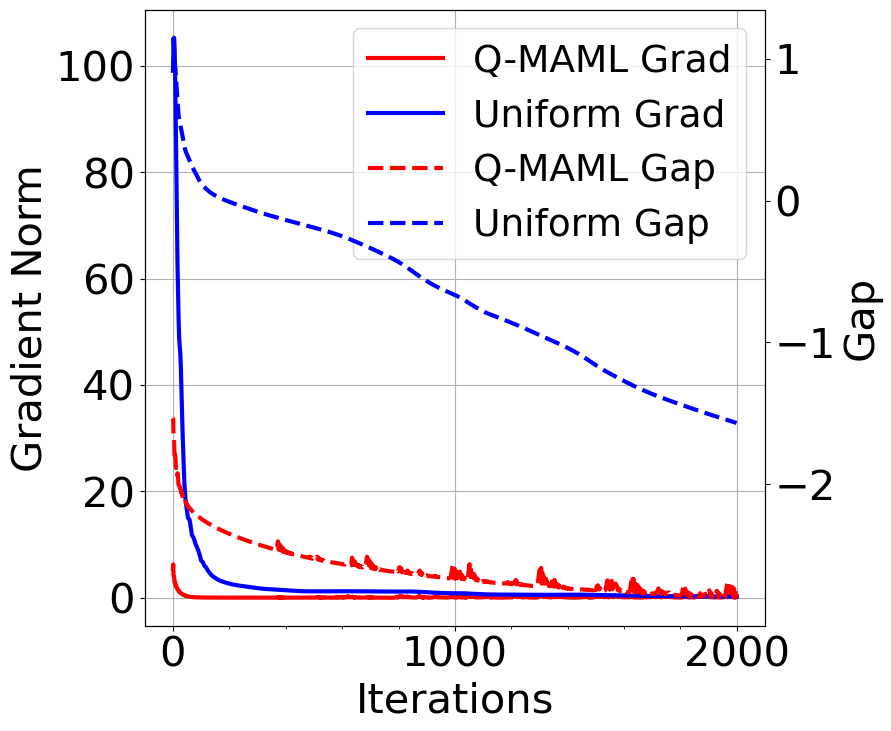} \label{fig:disc_xyz_vqa_grad}}
     \subfigure[Molecule 10 qubits]{\includegraphics[width=0.47\columnwidth]{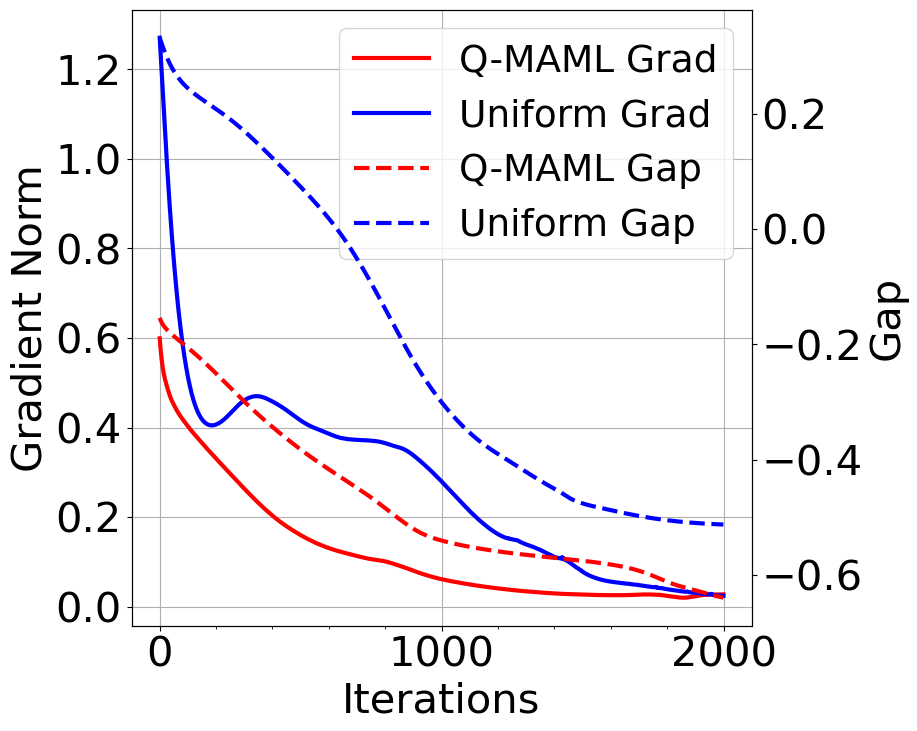} \label{fig:mol_vqa_grad}}
    \caption{Fluctuation of gradient norm of PQC while training in the adaptation phase.}
    \label{fig:vqa_grad}
\end{figure}

To further investigate the challenge of barren plateaus, we focus on the adaptation process and use gradient norms as a diagnostic tool. Examining the gradient norms throughout the adaptation process allows us to identify the effective regions of the learning curve where optimization is most effective. In addition, we analyze the statistical properties of these regions to assess the suitability of the Q-MAML initialization and adaptation strategy. This approach provides a deeper understanding of how our method navigates these landscapes, ultimately enhancing the efficiency and effectiveness of the optimization process.

During the adaptation process, the relative positions of each initial parameter can be empirically compared using the gradient norm and the gap between the expected value and ground state energy. Based on Figure~\ref{fig:grad_curve} and \ref{fig:vqa_grad}, the parameters initialized by Q-MAML exhibit a moderate gradient norm and are positioned close to the optimal solution, allowing for quicker convergence. Additionally, the mean and variance of these parameters differ from those produced by traditional methods. 

\begin{figure}[t]
    \centering
    \includegraphics[width=0.9\linewidth]{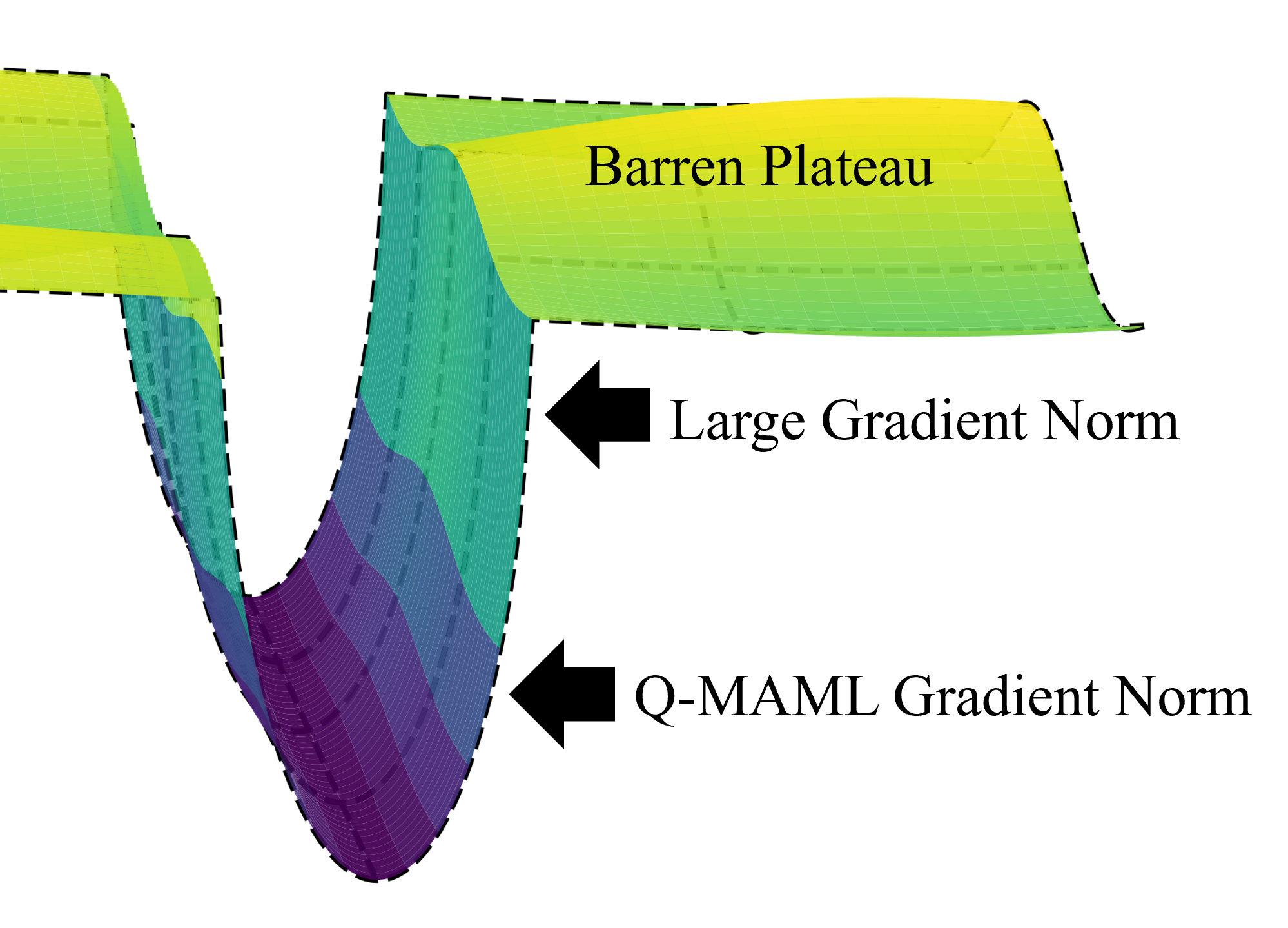}
    \caption{Visualization of optimization landscape in Q-MAML. The "Barren Plateau" region, is depicted in yellow. The green and blue regions indicate areas with "Large Gradient Norm" and "Q-MAML Gradient Norm".} 
    \label{fig:discussion_otland}
\end{figure}

Building on these observations, we provide a conceptual figure of optimization landscape in~\reffig{discussion_otland}. Through observation, we can imply that the initialization point provided by Q-MAML is located in space with a moderate gradient norm, which is closer to the optimal value. On the contrary, slower convergence, and a larger gradient norm of Uniform initialization can be understood as located far from the optimal value.

Similar to the context of classical neural networks training, excessively large gradients can cause unstable updates to the parameters which leads to instability and slow convergence.
Our findings suggest that the initialization points identified by Q-MAML reside in regions of the parameter space where gradient norms remain moderate, effectively mitigating the problem of gradients. 



\section{Conclusion} \label{sec:con}
We introduced a quantum-classical hybrid framework for optimizing PQCs utilizing a MAML-inspired approach. Our methodology aims to improve the initial parameter of PQC to enhance convergence speed in the adaptation phase, which is critical for effective quantum computation in the NISQ era. Through rigorous experimentation, our results demonstrate that Q-MAML significantly improves the training dynamics of PQCs across various quantum task spaces, including Heisenberg XYZ Hamiltonian and Molecule Hamiltonian.

Our analyses reveal that Q-MAML effectively circumvents the common challenges posed by barren plateaus through superior initialization strategies, which are tailored to specific quantum tasks by learning from a broad task landscape. This approach not only facilitates a deeper understanding of the gradient norms but also allows us to pinpoint optimal learning regions, thus contributing to more efficient and precise quantum optimization.

Furthermore, the comparative study against traditional initialization methods like Uniform and Gaussian showed that Q-MAML’s initialization leads to better positioning of the learning trajectory, enhancing the model's adaptability and overall performance. The implementation of this framework holds great promise for advancing the capabilities of quantum computing algorithms, particularly in solving complex optimization problems that are pivotal in the field of quantum mechanics.

Building on our findings, we raise the issue of "optimal gradients" in VQAs, a challenge similar to that faced in classical neural networks. Our future work will focus on refining the Q-MAML framework to incorporate more diverse quantum tasks and explore further applications in other quantum computing paradigms. We believe that the continued development of meta-learning strategies in quantum computing will significantly contribute to overcoming the current limitations of PQCs and pave the way for more robust, scalable quantum technologies.

\section{Acknowledgements}
This research was supported by Quantum Computing based on Quantum Advantage challenge research(RS-2023-00257561) through the National Research Foundation of Korea(NRF) funded by the Korean government (Ministry of Science and ICT(MSIT)).
\bibstyle{aaai25}
\bibliography{qmaml_main}

\newpage

\end{document}